\documentclass{article}

\usepackage{latexsym}

\usepackage{amssymb}

\newcommand{\R}{\mathbb{R}}
\newcommand{\tmax}{T_\mathrm{max}}
\newcommand{\supp}{\mathrm{supp}\,}
\newcommand{\N}{\mathbb{N}}
\def\prfe{\hspace*{\fill} $\Box$

\smallskip \noindent}
\newtheorem{Theorem}{Theorem}
\newtheorem{Proposition}{Proposition}
\newtheorem{Lemma}{Lemma}
\newtheorem{Corollary}{Corollary}

\title{On classical solutions \\of the Nordstr\"om-Vlasov system}

\author
{ Simone Calogero \& Gerhard Rein\\
Institut f\"ur Mathematik der
Universit\"at Wien\\
Strudlhofgasse 4\\
1090 Vienna, Austria}
\date { }
                                                                             
\begin{document}
\maketitle 
\begin{abstract}
The Nordstr\"om-Vlasov system describes the dynamics of a self-grav\-it\-ating 
ensemble of collisionless particles
in the framework of the Nordstr\"om scalar theory of gravitation. 
We prove existence and uniqueness of classical 
solutions of the Cauchy problem in three dimensions and establish a 
condition which guarantees that the solution is 
global in time. Moreover, we show that if one changes the sign 
of the source term in the field 
equation, which heuristically corresponds to 
the case of a repulsive gravitational force, then solutions
blow up in finite time for a large class of initial data.
Finally, we prove global existence of classical solutions
for the one dimensional version of the system 
with the correct sign in the field equation. 
\end{abstract}

\section{Introduction} \label{intro}
\setcounter{equation}{0}
In astrophysics, systems such as galaxies or globular clusters
are often modeled as a large ensemble of particles (stars) which
interact only by the gravitational field which they create
collectively. In such systems collisions among the particles are sufficiently
rare to be neglected. Let $\mathfrak{f}=\mathfrak{f}(t,x,p)\geq 0$ denote 
the density of the particles in phase-space, 
where $t\in\R$ denotes time, $x\in\R^3$
position, and $p\in\R^3$ momentum. This density function satisfies 
the Vlasov equation---a continuity equation on phase space---coupled
to the field equation(s) for the gravitational field.
In the non-relativistic case this results in the Vlasov-Poisson
system, in the relativistic setting one obtains the Einstein-Vlasov system. 
While for the former system the initial value problem is by now
well understood---cf.\ \cite{LP,Pf,R,Sch}---this is not so
for the much more 
difficult Einstein-Vlasov system, the difficulty arising from the different
nature of the field equations. We refer to \cite{Ha} for a recent review on 
the latter system, cf.\ also \cite{RR,Rl}.
In the present paper we investigate a different relativistic model 
which is obtained by coupling the Vlasov equation to the
Nordstr\"om scalar gravitation theory \cite{No}. In this theory, the 
gravitational effects are mediated by a scalar field $\phi$, and the system 
reads  
\begin{equation} \label{wave1}
\partial_t^2\phi-\bigtriangleup_x\phi=-e^{4\phi}
\int\mathfrak{f}\,\frac{dp}{\sqrt{1+p^{2}}},
\end{equation}
\begin{equation} \label{vlasov1}
\partial_{t}\mathfrak{f} + \widehat{p}\cdot\nabla_x\mathfrak{f} -
\left[\left(\partial_{t}\phi + \widehat{p}\cdot\nabla_x\phi \right) p +
(1+p^{2})^{-1/2}\nabla_x\phi\right]\cdot\nabla_p\mathfrak{f}=0.
\end{equation}
Here  
\[
\widehat{p} =\frac{p}{\sqrt{1+p^2}}
\]
denotes the relativistic velocity of a particle with momentum $p$,
$p^2 = |p|^2$,
and units are chosen such that the mass of each particle, 
the gravitational constant and the speed of light are 
all equal to unity.
A solution $(\mathfrak{f},\phi)$ of this system 
is interpreted as follows: The space-time is a 
Lorentzian manifold with a conformally flat metric which, in the coordinates 
$(t,x)$, takes the form
\[
g_{\mu\nu}=e^{2\phi} \textrm{diag}(-1,1,1,1).
\]
The particle distribution defined on the mass shell
in this metric is given by
\begin{equation} \label{fph}
\mathfrak{f}_\mathrm{ph}(t,x,p)=\mathfrak{f}(t,x,e^\phi p).
\end{equation}
More details on the derivation of this model are given in \cite{C}.
It should be emphasized that the system does not constitute a
physically correct model. Our interest in it is motivated by the
fact that while being much simpler than the ``correct'' relativistic
model, the Einstein-Vlasov system, it still captures some of the
essential features which distinguish the latter from non-relativistic
models. In particular, the Nordstr\"om-Vlasov model
does allow for propagation of gravitational waves, 
even in the spherically symmetric
case. The hope is that the analysis
of this model will lead to a better mathematical understanding
of a whole class of nonlinear partial differential equations
and eventually to a better understanding of the Einstein-Vlasov
system. However, scalar fields {\em do} play a major role in modern 
theories of (classical and quantum) gravity,
cf. \cite{DE} and references therein, as well as in numerical 
investigations of relativistic gravitational waves, \cite{ST}.

It turns out to be convenient to rewrite the system
in terms of the new unknowns $(f,\phi)$, where $f$ is given by
\begin{equation} \label{ourf}
f(t,x,p)=e^{4\phi(t,x)} \mathfrak{f}(t,x,p).
\end{equation}
The system then takes the form
\begin{equation}\label{wave2}
\partial_t^2\phi-\bigtriangleup_x\phi=-\mu,
\end{equation}
\begin{equation} \label{mudef}
\mu(t,x) = \int f(t,x,p)\,\frac{dp}{\sqrt{1+p^2}},
\end{equation}
\begin{equation} \label{vlasov2}
Sf
- \left[(S\phi)\,p + (1+p^2)^{-1/2} \nabla_x\phi \right]\cdot\nabla_p f
= 4 f\, S\phi\,,
\end{equation}
where
\[
\quad S =\partial_t+\widehat{p}\cdot\nabla_x
\]
is the free-transport operator.
The function $\mu$
is the trace of the energy momentum tensor.
In the following we refer to the system (\ref{wave2}), (\ref{mudef}),
(\ref{vlasov2}) as the Nordstr\"om-Vlasov system.
We supply it with the initial conditions
\begin{equation} \label{data}
f(0,x,p)=f^\mathrm{in}(x,p),\quad \phi(0,x)=\phi_0^\mathrm{in}(x),
\quad\partial_t
\phi(0,x)=\phi_1^\mathrm{in}(x),\quad x,\;p \in \R^3.
\end{equation}
A system of equations which belongs to the same class of kinetic
models and is used in plasma physics is the relativistic Vlasov-Maxwell system.
For this system Glassey and Strauss proved that classical solutions
to the initial value problem with regular data exist 
and no shocks or singularities form as long as the maximal momentum
of the particles is controlled, cf.\ \cite{GS}.
One purpose of the present paper is to establish an analogous
condition for global existence of classical solutions for the
Nordstr\"om-Vlasov system.
In order to state the result in a precise way, let us consider the quantity
\[
\mathcal{P}(t)=\sup \{|p|: 0\leq s< t,\ (x,p) \in \supp f(s)\},
\]
the maximum momentum of the particles up to the time $t$;
$f(s)$ will always be compactly supported.
In addition, we need to monitor the quantity
\[
\mathcal{Q}(t)=\sup\{|\phi(s,x)|: \, 0\leq s <t,\,(x,p)\in\supp f(s)\},
\]
and we define $\Lambda(t)=\mathcal{P}(t)+\mathcal{Q}(t)$;
for $t=0$ we define these quantities by what is induced by the initial data.
\begin{Theorem} \label{locex}
Initial data $f^\mathrm{in}\in C_c^1(\R^6)$, 
$\phi_0^\mathrm{in}\in C_b^3 (\R^3)$, $\phi_1^\mathrm{in}\in C_b^2 (\R^3)$
with $f^\mathrm{in}\geq 0$
launch a unique classical  solution 
$(f,\phi) \in C^1([0,\tmax[\times\R^6) 
\times C^{2}([0,\tmax[\times\R^3)$ 
of the Nordstr\"om-Vlasov system (\ref{wave2}), (\ref{mudef}), (\ref{vlasov2}) 
in a maximal interval of time,
satisfying the initial conditions (\ref{data}).
If $\Lambda(\tmax)<\infty$
then $\tmax=\infty$, i.e., the solution $(f,\phi)$ is global in time.
\end{Theorem}
We restrict ourselves to study the Cauchy problem forward 
in time; the system is time-reversible.
The notation on the function spaces above is standard:
The index $c$ refers to compactly supported functions, and the
index $b$ means that all derivatives up to the indicated order are
bounded. This boundedness condition which only aims to make our 
estimates more transparent can be removed using the 
finite propagation speed which in turn 
follows from local conservation of energy, see below. 
We observe that unlike the case of \cite{GS} where
global existence follows from a bound on 
$\mathcal{P}(t)$ alone we require a bound on $\phi$ 
in the support of $f$ in addition. 
This will imply a bound on $f$; 
as opposed to the Vlasov-Maxwell system this
bound on $f$ is a priori not obvious since $f$ is not constant 
along characteristics of the Vlasov equation.    
Note however that by (\ref{fph}) the ``physical'' momentum of the 
particles is obtained by rescaling $p$ by the factor $e^{\phi}$ so that
the meaning of Theorem~\ref{locex} remains the same as with the analogous 
result for the Vlasov-Maxwell system: 
The occurrence of singularities in the Nordstr\"om-Vlasov 
system---if any---can only be due to particles which pick up arbitrarily 
large momentum as the system evolves. 

The proof of Theorem~\ref{locex} is presented in 
Sections~\ref{prel}--\ref{locexproof}, a
key ingredient being certain 
representation formulas for the first and second order derivatives of $\phi$ 
which we give in Section~\ref{fieldrep}. These formulas are 
similar to the ones introduced by Glassey and Strauss for the 
relativistic Vlasov-Maxwell
system, and we will be able to reduce many details to the 
case of \cite{GS}. 

An interesting mathematical feature of the Nordstr\"om-Vlasov
system is that if we change the sign in the right 
hand side of (\ref{wave2}) then solutions blow up in finite
time for a large class of initial data:
\begin{Theorem} \label{blowup}
Consider the system  obtained by coupling (\ref{mudef}), (\ref{vlasov2}) to the 
equation
\[
\partial_t^2\phi-\bigtriangleup_x\phi=\mu
\]
instead of (\ref{wave2}).
Then the assertion of Theorem~\ref{locex} remains valid.
If the initial data are homogeneous (i.e. 
independent of $x$) in a ball $B_R(0)=\{x\in\R^3:|x|\leq R\}$
for some $R>0$ and if 
\[
R^2 \mu(0,0)\geq 2,\quad \phi_1^\mathrm{in}(0)\geq\sqrt{2 \mu(0,0)},
\]
then $\tmax\leq R$.
\end{Theorem}
This result will be shown in Section~\ref{blowupproof}.
It is interesting because the blow-up results from the nonlinearity
in the original field equation, which has only been hidden
by rewriting the system in the form (\ref{wave2}), (\ref{mudef}),
(\ref{vlasov2}),
and which is a feature that distinguishes this system from the
relativistic Vlasov-Maxwell system where the field equations by themselves
are linear. 

Solutions of the Nordstr\"om-Vlasov system preserve energy,
which is positive definite in the situation of Theorem~\ref{locex}
but indefinite in the situation of Theorem~\ref{blowup}.
For the three dimensional system considered so far we are not
going make use of conservation of energy. However, in the last section
we will show that conservation of energy can be used to prove
global existence of classical solutions to the
one dimensional version of the system: 
\begin{equation}\label{wave1d}
\partial_t^2\phi-\partial_x^2\phi=-\mu,
\end{equation}
\begin{equation} \label{mudef1d}
\mu(t,x) = \int f(t,x,p)\,\frac{dp}{\sqrt{1+p^2}},
\end{equation}
\begin{equation} \label{vlasov1d}
S f
- \left[ (S\phi)\,p + (1+p^2)^{-1/2} \partial_x\phi \right] \partial_p f
= 2 f\,S\phi,
\end{equation}
where now $x,p \in \R$.
The right hand side of the Vlasov equation  modified 
so as to retain conservation of energy, cf.\ the next section.
Again, the sign in the field equation
turns out to be important. While the blow-up result of Theorem~\ref{blowup}
also holds in other dimensions we
have the following global existence result for the ``correct'' sign:
\begin{Theorem} \label{globex1d}
Initial data $f^\mathrm{in}\in C_c^1(\R^2)$, 
$\phi_0^\mathrm{in}\in C_b^2 (\R)$, $\phi_1^\mathrm{in}\in C_b^1 (\R)$
with $f^\mathrm{in}\geq 0$ launch a unique global classical  solution 
$(f,\phi) \in C^1([0,\infty[\times\R^2) 
\times C^2([0,\infty[\times\R)$ 
of the system (\ref{wave1d}), (\ref{mudef1d}), (\ref{vlasov1d}).
\end{Theorem}
\section{General properties of the Nordstr\"om-Vlasov system in $N$ 
         spatial dimensions} \label{general}
\setcounter{equation}{0}
In this section we collect some general properties of the system
under consideration. Some of these will be important in the sequel
while others are included for reasons of general interest. Since in 
Section \ref{1d} we consider the one dimensional version of the
Nordstr\"om-Vlasov system, we discuss here the generalization 
to the case of $N$ spatial dimensions, $N\geq 1$.  
This generalization is defined as the
system obtained by adapting the derivation of \cite{C} to the case in 
which the space-time is a $1+N$ dimensional Lorentzian manifold. This
assumption affects the argument of \cite{C} in a non-trivial way only 
in two points. Firstly, the modulus of the determinant of the physical
metric, denoted by $\mathfrak{g}$ in \cite{C}, changes into 
$\mathfrak{g}=\exp[2(1+N)\phi)]$. Secondly, the rescaling law which 
relates the stress-energy tensor in the physical and unphysical frames 
becomes, using the notation of \cite{C}, $T_*^{\mu\nu}=\exp[(N+3)\phi]T^{\mu\nu}$.
The latter identity follows by the definition of the stress-energy tensor,
which is independent of the frame and of the dimension, as the
variation with respect to the metric of the matter action. 
With these modifications one can show that the generalization of 
(\ref{wave2}), (\ref{mudef}), (\ref{vlasov2}) in $1+N$ dimensions is obtained by taking $t\in\R$, $x\in\R^N$, $p\in\R^N$ and replacing 
the Vlasov equation (\ref{vlasov2}) by 
\begin{equation} \label{vlasov2n}
Sf
- \left[(S\phi)\,p + (1+p^2)^{-1/2} \nabla_x\phi \right]\cdot\nabla_p f
= (1+N) f\, S\phi\,.
\end{equation}
The resulting system reduces to (\ref{wave2}), (\ref{mudef}), (\ref{vlasov2}) in the case $N=3$ and to (\ref{wave1d}), (\ref{mudef1d}), 
(\ref{vlasov1d}) for $N=1$.

An important tool in the sequel are the characteristics of the
Vlasov equation. Let $(X,P)(s)=(X,P)(s,t,x,p)$ denote the solution of
\begin{eqnarray}
\frac{dx}{ds}
&=&
\frac{p}{\sqrt{1+p^2}}, \label{xchar}\\
\frac{dp}{ds}
&=&
-(S\phi)(s,x,p)\,p -\frac{\nabla_x\phi(s,x)}{\sqrt{1+p^2}},
\label{pchar}
\end{eqnarray}
with initial condition
\begin{equation} \label{chardata}
(X,P)(t,t,x,p)=(x,p).
\end{equation}
Integrating (\ref{vlasov2n}) along characteristics we obtain
\begin{equation} \label{frep1}
f(t,x,p)=f^\mathrm{in}(X(0),P(0)) +
(1+N)\int_0^t (f\, S \phi)(\tau,X(\tau),P(\tau))\,d\tau.
\end{equation}
Moreover (\ref{vlasov2n}) can be rewritten as
\begin{equation} \label{vlasov3n}
S(e^{-(1+N)\phi}f)
- \left[(S\phi)\,p + (1+p^2)^{-1/2} \nabla_x\phi \right]\cdot\nabla_p (e^{-(1+N)\phi}f)
= 0,
\end{equation}
showing that the function 
$e^{-(1+N)\phi}f$ is constant along solutions of (\ref{xchar}), (\ref{pchar})
(which of course are the characteristics of (\ref{vlasov3n}) as well), 
and hence the function $f$ can be also represented as follows:
\begin{equation} \label{frep2}
f(t,x,p)=f^\mathrm{in}(X(0),P(0))
\exp \left[(1+N)\phi(t,x)-(1+N)\phi^\mathrm{in}(X(0))\right].
\end{equation}
By (\ref{frep1}), the function 
$\zeta(s)=f(s,X(s),P(s))$ solves the differential equation
$\dot{\zeta}=(1+N)\,\dot{\eta}(s)\,\zeta$, with $\eta(s)=\phi(s,X(s))$. 
Integration of this equation yields (\ref{frep2}).
The latter equation shows that the compact support property
of $f(t)$ which is assumed initially propagates. 

It should be noted that the characteristic flow is not measure preserving
on $\R^{2N}$. Indeed, the fact that the $(x,p)$-divergence 
of the right hand
side of (\ref{xchar}), (\ref{pchar}) equals $- N\, S\phi$ implies that
\[
\det \frac{\partial(X,P)}{\partial(x,p)}(0,t,x,p) = 
\exp\left[N\,\phi(t,x) - N\,\phi^\mathrm{in}(X(0,t,x,p))\right], 
\]
which together with (\ref{frep2}) implies 
conservation of rest-mass:
\[
\int\!\!\!\int  f(t,x,p)e^{-\phi(t,x)}\,dx\, dp =\textrm{const}.
\] 
The Nordstr\"om-Vlasov system also satisfies an energy conservation law. 
In the formulation 
of the system used here, the energy density is defined by
\[
e(t,x)=\int \sqrt{1+p^2}f(t,x,p)\, dp + \frac{1}{2}(\partial_t\phi)^2+
\frac{1}{2}(\nabla_x\phi)^2,
\]
and the momentum density by
\[
\mathfrak{p}(t,x)=\int p\, f(t,x,p)\,dp -\partial_t\phi\nabla_x\phi.
\]
It is easy to check that local energy conservation holds,
\[
\partial_t\,e + \nabla_x\cdot\mathfrak{p}=0,
\]
which upon integration leads to conservation of the total 
energy:
\[
\int\!\!\!\int \sqrt{1+p^2} f(t,x,p)\,dp\,dx +
\frac{1}{2}\int \big [(\partial_t\phi)^2+
(\nabla_x\phi)^2\big](t,x)\,dx =\textrm{const.}
\]
We note that the definition of energy and rest mass is independent of the 
dimension.
Moreover, the total energy is positive definite. 
However, the sign in front of the field energy is reversed
for the ``repulsive'' case considered in Theorem~\ref{blowup},
and so the total energy is no longer positive definite in that case.

The contents of this section will be applied for $N=1$ in Section \ref{1d} 
and for $N=3$ in the rest of the paper
without further comment.
\section{Preliminary estimates} \label{prel}
\setcounter{equation}{0}

We fix some notation. The usual $L^\infty$-norm is denoted
by $\|\cdot\|_\infty$, $\|\cdot\|_{k,\infty}$ is the $L^\infty$-norm
of the argument and its derivatives up to order $k$. Moreover,
\begin{eqnarray*}
\|D f(t)\|_\infty
&=&
\sup\{|\partial_{x_i} f(t,x,p)|,\; |\partial_{p_i} f(t,x,p)| :
(x,p)\in\R^6,\ i=1,2,3\},\\
\|D\phi(t)\|_\infty
&=&
\sup\{|\partial_t\phi(t,x)|,\; |\partial_{x_i}\phi(t,x)| :
\,x\in\R^3,\ i=1,2,3\},\\
\|D^2 \phi(t)\|_\infty
&=&
\sup\{|\partial^2_t\phi(t,x)|,\; |\partial_t\partial_{x_i}\phi(t,x)|,\;
|\partial_{x_i}\partial_{x_j}\phi(t,x)| :
\,x\in\R^3,\ i,j = 1,2,3\}.
\end{eqnarray*}
We shall use similar definitions for the one dimensional case in Section \ref{1d}.
Numerical constants are denoted by $c$ and may change from line to line. 
Constants depending on the initial data are denoted by $C,
C_0, C_1,...$, dependence on the length of some time
interval is denoted by $C_T$. 
Since $f^\mathrm{in}$ has compact support
\[
P_0=\sup\{|p|: (x,p) \in \supp f^\mathrm{in}\} < \infty;
\]
without loss of generality we may assume that $P_0\geq 1$
and hence $\mathcal{P}(t)\geq 1$ for all $t$, otherwise
we add 1 to both quantities.

We prove a few preliminary estimates which will be used in the 
proof of Theorem~\ref{locex}. By solution we always mean a solution
of the Nordstr\"om-Vlasov system (\ref{wave2}), (\ref{mudef}), (\ref{vlasov2})
with the regularity and initial
data as specified in Theorem~\ref{locex}.
\begin{Lemma} \label{fest}
For any solution defined on some interval $[0,T[$, 
\[
\|f(t)\|_\infty\leq \|f^\mathrm{in}\|_\infty+4\int_0^t 
\|D \phi(\tau)\|_\infty \|f(\tau)\|_\infty d\tau,\ t\in [0,T[.
\]
If $\mathcal{Q}(T)<\infty$ then  
\[
\sup\{\|f(t)\|_\infty:\, 0\leq t<T\}<\infty.
\]  
\end{Lemma}
\noindent\textit{Proof: }
The estimates are immediate by (\ref{frep1}) and (\ref{frep2}). \prfe

\begin{Lemma} \label{phiest}
For any solution defined on some interval $[0,T[$,
\[
\|\phi(t)\|_\infty\leq (1+t)\left[C_0 + c \int_0^t \|f(\tau)\|_\infty
\mathcal{P}(\tau)^{2}\,d\tau \right],\ t \in [0,T[,
\]
where $C_0=c\,(\|\phi^\mathrm{in}_0\|_{1,\infty}+ 
\|\phi^\mathrm{in}_1\|_{\infty})$. If $\Lambda(T)<\infty$ then
\[
\sup\{\|\phi(t)\|_\infty:\,0\leq t< T\}<\infty,
\]
i.e., $\phi$ is bounded everywhere and not just on
the $x$-support of $f$.

\end{Lemma}
\noindent\textit{Proof: }By Kirchhoff's formula,
\begin{equation} \label{kirchhoff}
\phi(t,x) = \phi_0(t,x) 
- \int_{|x-y|\leq t}
  \int f(t-|x-y|,y,p)\frac{dp}{\sqrt{1+p^2}}\frac{dy}{|x-y|}.
\end{equation}
The first term is the solution of the homogeneous wave 
equation with data $(\phi^\mathrm{in}_0,\phi^\mathrm{in}_1)$ and is 
estimated by
\[
\|\phi_0(t)\|_\infty \leq c (1+t)(\|\phi^\mathrm{in}_0\|_{1,\infty}+
\|\phi^\mathrm{in}_1\|_{\infty})=C_0(1+t).
\]
The second term is estimated by
\begin{eqnarray*}
&&\int_0^t\int_{|x-y|=t-\tau}
\int_{|p|\leq \mathcal{P}(\tau)} f(\tau,y,p)
\frac{dp}{\sqrt{1+p^2}} dS_y \frac{d\tau}{t-\tau}\\
&&\qquad
\leq c \int_0^t (t-\tau)\|f(\tau)\|_\infty \mathcal{P}(\tau)^{2}\,d\tau 
\leq c (1+t)\int_0^t \|f(\tau)\|_\infty \mathcal{P}(\tau)^{2}\,d\tau.
\end{eqnarray*}
Together with Lemma~\ref{fest}
this completes the proof. \prfe

\begin{Lemma} \label{pest}
The function $\mathcal{P}$ satisfies the estimate
\[
\mathcal{P}(t)\leq P_0 + 2\int_0^t \|D \phi(\tau)\|_\infty
(1+\mathcal{P}(\tau))
\,d\tau.
\]
\end{Lemma}
\noindent\textit{Proof: }This follows by integrating (\ref{pchar}) 
and observing that the right hand side 
\[
- (\partial_t\phi (\tau,x) + \widehat{p}\cdot \nabla_x  \phi(\tau,x)) p -
(1+p^2)^{-1/2}\nabla_x \phi(\tau,x) 
\]
of that equation is estimated by $2 \|D \phi(\tau)\|_\infty (1+|p|)$. \prfe

To conclude this section we estimate the derivatives of the 
characteristics with respect to their initial data
in terms of the derivatives of $\phi$. 
\begin{Lemma} \label{dcharest}
Assume that $\mathcal{P}(T)<\infty$. Then 
for all $t \in [0,T[$ and $x,p \in \R^3$,
\[
|\nabla_{(x,p)}(X,P)(0,t,x,p)|\leq C_T\exp\left[C_T\int_0^t
\left(1+\|D \phi(\tau)\|_\infty + \|D^2\phi(\tau)\|_\infty\right)\,d\tau \right].
\]
\end{Lemma}
\noindent\textit{Proof: }Differentiating 
(\ref{xchar}) and (\ref{pchar}) with respect to $x_i$ and integrating
we obtain
\begin{eqnarray*}
\partial_{x_i} X_j(s)
&=&
\delta_{ij}+\int_t^s \frac{\partial \widehat{p}_k}{\partial p_j}(P(\tau))
\partial_{x_i}P_k(\tau)\,d\tau,\\
\partial_{x_i} P_j(s)
&=&
-\int_t^s \left[(S\partial_{x_k}\phi)\,p_j + (1+p^2)^{-1/2} \partial_{x_k}
\partial_{x_j}\phi \right](\tau,X(\tau),P(\tau))\partial_{x_i} X_k(\tau)\,d\tau\\
&&
{}-\int_t^s
\left[(S \phi)\,\delta_{jk}+\frac{\partial \widehat{p}_k}{\partial p_j}
p_l\partial_{x_l}\phi -
\frac{\widehat{p}_k\partial_{x_j}\phi}{1+p^2}\right](\tau,X(\tau),P(\tau))
\partial_{x_i}P_k(\tau)\,d\tau,
\end{eqnarray*}
summation over repeated indexes being understood. Now
\[
\frac{\partial \widehat{p}_k}{\partial p_j} =
\frac{(1+p^2)\delta_{jk}-p_j p_k}{(1+p^2)^{3/2}} .
\]
By assumption, $P(\tau)\leq C_T$, and so we get, for 
$s\leq t$,
\begin{eqnarray*}
|\nabla_x X(s)|
&\leq& 
c+\int_s^t |\nabla_x P(\tau)|\,d\tau,\\
|\nabla_x P(s)|
&\leq& 
C_T\int_s^t \bigl(\|D\phi(\tau)\|_\infty +
\|D^2 \phi(\tau)\|_\infty\bigr)
(|\nabla_x X(\tau)|+|\nabla_x P(\tau)|)\,d\tau.
\end{eqnarray*}
If we add these estimates and use Gronwall's
inequality we obtain the assertion for the $x$-derivatives,
and the argument for the $p$-derivatives is analogous. \prfe

\section{Representation and estimates for $D\phi$ and $D^2\phi$} \label{fieldrep}
\setcounter{equation}{0}
Let us first recall the structure of the integral representation for the 
electromagnetic field $(E,B)$ in the case of the relativistic 
Vlasov-Maxwell system, cf.\ \cite[Thm.~3]{GS}:
\begin{eqnarray}
E(t,x)
&=&
E_D(t,x)-\int_{|x-y|\leq t}\int a^E(\omega,p)
f(t-|x-y|,y,p)\,dp\,\frac{dy}{|x-y|^2}\nonumber\\
&&
{} -\int_{|x-y|\leq t}\int  \tilde{a}^E(\omega,p) 
(S f) (t-|x-y|,y,p)\,dp\,\frac{dy}{|x-y|},\label{erep}\\
B(t,x)
&=&
B_D(t,x)-\int_{|x-y|\leq t}\int a^B(\omega,p)
f(t-|x-y|,y,p)\,dp\,\frac{dy}{|x-y|^2}\nonumber\\
&&
{} -\int_{|x-y|\leq t}\int  \tilde{a}^B(\omega,p) 
(S f) (t-|x-y|,y,p)\,dp\,\frac{dy}{|x-y|}. \label{brep}
\end{eqnarray}
Here $\omega=(y-x)/|x-y|$, $E_D, B_D$ are integrals of 
the initial data and the kernels are given by
\begin{eqnarray*}
&&a^E(\omega,p)=\frac{\omega+\widehat{p}}{(1+p^2)
(1+\omega\cdot\widehat{p})^2},
\quad 
\tilde{a}^E(\omega,p)=\frac{\omega+\widehat{p}}{1+\omega\cdot\widehat{p}},\\
&&a^B(\omega,p)=\frac{\omega\wedge\widehat{p}}{(1+p^2)
(1+\omega\cdot\widehat{p})^2},\quad
\tilde{a}^B(\omega,p)=\frac{\omega\wedge\widehat{p}}{1+\omega\cdot\widehat{p}}.
\end{eqnarray*}
In the next two propositions we show that the derivatives of $\phi$ 
satisfy very similar integral representations.
\begin{Proposition} \label{dtphirep}
The time derivative of $\phi$ satisfies the representation formula
\begin{eqnarray*}
\partial_t\phi(t,x)
&=&
\partial_t\phi_0(t,x)-t^{-1}\int_{|x-y|=t}
\int \frac{f^\mathrm{in}(y,p)}{(1+\omega\cdot\widehat{p})}
\frac{dp}{\sqrt{1+p^2}} dS_y\\
&&
{} -\int_{|x-y|\leq t}\int  a^{\phi_t}
(\omega,p)f(t-|x-y|,y,p)\,dp\frac{dy}{|x-y|^2}\\
&&
{} -\int_{|x-y|\leq t}\int  b^{\phi_t}
(\omega,p)f\,(S\phi)(t-|x-y|,y,p)\,dp \frac{dy}{|x-y|}\\
&&
{} -\int_{|x-y|\leq t}\int  c^{\phi_t}
(\omega,p)\cdot\nabla_x\phi f(t-|x-y|,y,p)\,dp\frac{dy}{|x-y|},
\end{eqnarray*}
where the kernels are given by
\begin{eqnarray*}
a^{\phi_t}(\omega,p)
&=&
-\frac{(\omega+\widehat{p})\cdot\widehat{p}}
{\sqrt{1+p^2}(1+\omega\cdot\widehat{p})^2}
=-a^E(\omega,p)\cdot p,\\
b^{\phi_t}(\omega,p)
&=&
\frac{(\omega+\widehat{p})^2}
{\sqrt{1+p^2}(1+\omega\cdot\widehat{p})^2},\\
c^{\phi_t}(\omega,p)
&=&
\frac{\omega+\widehat{p}}{(1+p^2)^{3/2}
(1+\omega\cdot\widehat{p})^2}.
\end{eqnarray*}
They satisfy the estimates
\[
|a^{\phi_t}(\omega,p)|\leq c (1+p^2),\quad 
|b^{\phi_t}(\omega,p)|\leq c \sqrt{1+p^2},\quad 
|c^{\phi_t}(\omega,p)|\leq c.
\]
\end{Proposition}
\noindent\textit{Proof: }The claim is proved by a calculation 
similar to the one in \cite[Thm.~3]{GS}. We give only some of the details. 
By (\ref{kirchhoff}),
\begin{eqnarray*}
\partial_t\phi
&=&
\partial_t\phi_0-t^{-1}
\int_{|x-y|=t}\int f^\mathrm{in}(y,p)\frac{dp}{\sqrt{1+p^2}} dS_y\\
&&
{} -\int_{|x-y|\leq t}
\int\partial_t f(t-|x-y|,y,p)\frac{dp}{\sqrt{1+p^2}}\frac{dy}{|x-y|}.
\end{eqnarray*} 
In the last integral we use the identity
\begin{eqnarray}
(\partial_t g)(t-|x-y|,y,p)
= (1+\omega\cdot\widehat{p})^{-1}
&\Bigl[&
(S g)(t-|x-y|,y,p) \nonumber\\
&& -\widehat{p}\cdot\nabla_y [g(t-|x-y|,y,p)]\Bigr], \label{dtrep}
\end{eqnarray}
which holds for every $C^1$ function $g=g(t,x,p)$,
and integrate by parts in $y$ to obtain
\begin{eqnarray}
\partial_t\phi
&=&
\partial_t\phi_0(t,x)-t^{-1}
\int_{|x-y|=t}\int
\frac{f^\mathrm{in}(y,p)}{(1+\omega\cdot\widehat{p})}
\frac{dp}{\sqrt{1+p^2}}dS_y
\nonumber \\
&&
{}
-\int_{|x-y|\leq t}\int  
a^{\phi_t}(\omega,p)f(t-|x-y|,y,p)\,dp \frac{dy}{|x-y|^2} \nonumber\\
&&
{}
-\int_{|x-y|\leq t}\int \tilde{a}^{\phi_t}(\omega,p) 
(S f)(t-|x-y|,y,p)\,dp\frac{dy}{|x-y|},\label{dtphirep1}
\end{eqnarray}
where $\tilde{a}^{\phi_t}=(1+p^2)^{-\frac{1}{2}} 
(1+\omega\cdot\widehat{p})^{-1}$. 
Hence we see that the function
$\partial_t\phi$ has been written in the form (\ref{erep}) with $a^E$ 
replaced by $a^{\phi_t}$ and $\tilde{a}^E$ by
$\tilde{a}^{\phi_t}$.
To complete the proof we observe that by the Vlasov equation
(\ref{vlasov2}), 
$Sf=F(t,x,p)\cdot\nabla_p f+4f\,S \phi$
where $F(t,x,p)=(S\phi)\,p+(1+p^2)^{-1/2}\nabla_x\phi$
is the force term in the Vlasov equation. Substituting into the last 
integral of (\ref{dtphirep1}) and integrating by parts in $p$
the representation formula for $\partial_t \phi$ is 
obtained after a straightforward calculation. For the bounds on the
kernels we use the estimates
\begin{equation} \label{omegapest}
1+\omega\cdot\widehat{p}\geq\frac{1+|\omega\wedge p|^2}{2(1+p^2)},\quad 
|\omega+\widehat{p}|^2\leq 2(1+\omega\cdot\widehat{p}),
\end{equation}
cf.\ \cite[Lemma~1]{GS2}. Thus for example,
\[
|c^{\phi_t}(\omega,p)|\leq c (1+p^2)^{-3/2}
(1+\omega\cdot\widehat{p})^{-3/2}\leq
c (1+p^2)^{-3/2}\Big (\frac{1+p^2}{1+|p\wedge\omega|^2}\Big )^{3/2}\leq c,
\]
estimates for the other kernels being similar. \prfe
\begin{Proposition} \label{dxphirep}
The spatial derivatives of $\phi$ satisfy the representation formula
\begin{eqnarray*}
\partial_{x_i}\phi(t,x)
&=&
\partial_{x_i}\phi_0(t,x)-t^{-1}
\int_{|x-y|=t}\int
\frac{(1-\omega\cdot\widehat{p})}{1+\omega\cdot\widehat{p}}
\omega_i f^\mathrm{in}(y,p)\frac{dp}{\sqrt{1+p^2}}dS_y\\
&&
{}-\int_{|x-y|\leq t}\int  
a^{\phi_x}_i (\omega,p)f(t-|x-y|,y,p)\,dp\frac{dy}{|x-y|^2}\\
&&
{}-\int_{|x-y|\leq t}\int 
b^{\phi_x}_i (\omega,p)(f\,S \phi)(t-|x-y|,y,p)\,dp \frac{dy}{|x-y|}\\
&&
{}-\int_{|x-y|\leq t}\int  
c^{\phi_x}_i (\omega,p)\cdot\nabla_x\phi f(t-|x-y|,y,p)\,dp
\frac{dy}{|x-y|},
\end{eqnarray*}
where the kernels  are given by
\[
a^{\phi_x}=\sqrt{1+p^2}\,[a^E - \widehat{p}\wedge a^B],\quad 
b^{\phi_x}=b^{\phi_t} \omega,
\quad c^{\phi_x}_i=\omega_i c^{\phi_t},\ i=1,2,3.
\]
They satisfy the estimates
\[
|a^{\phi_x}(\omega,p)|\leq c (1+p^2),\quad|b^{\phi_x}
(\omega,p)|\leq c \sqrt{1+p^2},\quad  |c^{\phi_x}_i(\omega,p)|\leq c.
\]
\end{Proposition} 
\noindent\textit{Proof: } 
The only difference to the proof of Proposition~\ref{dtphirep}
is that now the identity
\begin{eqnarray} \label{dyrep}
(\partial_{y_{i}}g)(t-|x-y|,y,p)
&=&
\omega_{i}
(1+\omega\cdot\widehat{p})^{-1}(S g)(t-|x-y|,y,p) \nonumber\\
&&
{}+\Bigg(\delta_{ik}-\frac{\omega_{i}\widehat{p}_{k}}{1+\omega\cdot\widehat{p}}
\Bigg) \partial_{y_{k}}[g(t-|x-y|,y,p)]
\end{eqnarray}
is used instead of (\ref{dtrep}). 
Thus the analogue of (\ref{dtphirep1}) in this case is 
\begin{eqnarray*}
\nabla_x\phi(t,x)
&=&
(\nabla_x\phi)_D -\int_{|x-y|\leq t}
\int  a^{\phi_x}(\omega,p)f(t-|x-y|,y,p)\,dp \frac{dy}{|x-y|^2}\\
&&
{}-\int_{|x-y|\leq t}\int  \tilde{a}^{\phi_x}
(\omega,p) (S f)(t-|x-y|,y,p)\,dp \frac{dy}{|x-y|},
\end{eqnarray*}
where $(\nabla_x\phi)_D$ is the data term and 
$\tilde{a}^{\phi_x}=(1+p^2)^{-1/2} (1+\omega\cdot\widehat{p})^{-1}\omega$.
Since $a^E$ and $a^B$ are 
bounded by $c \sqrt{1+p^2}$ (cf.\ \cite{GS2}) 
the estimates on the kernels follow. \prfe 

We use the above representation formulas to estimate the first order
derivatives of $\phi$.
\begin{Lemma} \label{dphiest}
For any solution on some time interval $[0,T[$,
\[
\|D\phi (t)\|_\infty\leq (1+t)\left[C_1
+ c \int_0^t  
(1+\|D \phi(\tau)\|_\infty)\|f(\tau)\|_\infty
\mathcal{P}(\tau)^{5}\,d\tau \right],
\ t \in [0,T[,
\]
where $C_1=c \,P_0^4 (1+\|f^\mathrm{in}\|_\infty +
\|\phi^\mathrm{in}_0\|_{2,\infty}+\|\phi^\mathrm{in}_1\|_{1,\infty})$. 
If $\Lambda (T)<\infty$ then
\[
\sup\{\|D\phi (t)\|_\infty:\,0\leq t<T\}<\infty.
\]
\end{Lemma}
\noindent\textit{Proof: }By Proposition~\ref{dtphirep}, 
$\partial_t\phi$ is the sum of five 
terms which we denote by $I_1,\ldots,I_5$.
The first term is estimated as 
\[
|I_1|\leq c\, (\|\phi^\mathrm{in}_0\|_{2,\infty}+
\|\phi^\mathrm{in}_1\|_{1,\infty})(1+t).
\]
For the second term we use (\ref{omegapest}) to obtain
\[
|I_2|\leq c (1+t)\|f^\mathrm{in}\|_\infty\int_{|p|\leq P_0}\sqrt{1+p^2}\,dp
\leq c\,P_0^4 \|f^\mathrm{in}\|_\infty (1+t);
\]
recall that $P_0\geq 1$.
For $I_3$ we have, using the bounds on the kernels in 
Proposition~\ref{dtphirep},
\begin{eqnarray*}
|I_3|
&\leq&
c \int_0^t\mathcal{P}(\tau)^2
\int_{|x-y|=t-\tau}\int_{|p|\leq \mathcal{P}(\tau)}
f(\tau,y,p)\,dp\,dS_y\frac{d\tau}{(t-\tau)^2}\\
&\leq& 
c \int_0^t \|f(\tau)\|_\infty\mathcal{P}(\tau)^{5}\,d\tau.
\end{eqnarray*}
Likewise we obtain
\begin{eqnarray*}
|I_4|
&\leq& 
c\, (1+t)\int_0^t \|D \phi (t)\|_\infty
\|f(\tau)\|_\infty\mathcal{P}(\tau)^{4}\,d\tau,\\
|I_5|
&\leq& 
c\, (1+t)\int_0^t \|D \phi (t)\|_\infty\|f(\tau)\|_\infty
\mathcal{P}(\tau)^{3}\,d\tau.
\end{eqnarray*}
To complete the proof we add these estimates and 
repeat the same argument for $\nabla_x\phi$.
The bound on $D\phi$ follows from Gronwall's inequality together
with Lemma~\ref{fest}.\prfe

If we differentiate (\ref{frep2}) and use the estimates from
Lemmas~\ref{dcharest} and \ref{dphiest} we immediately obtain:
\begin{Corollary} \label{dfest}
Assume that $\Lambda(T)<\infty$. Then 
\[
\|D f(t)\|_\infty\leq C_{T}\exp\Big[C_T\int_0^t
(1+\|D^2 \phi(\tau)\|_\infty)\,d\tau \Big],\ t\in [0,T[.
\]
\end{Corollary}

Next we derive the integral representations and the 
necessary estimates for the second order derivatives of $\phi$.
\begin{Proposition} \label{ddtphirep}
The function $\partial_t^2\phi$ can be represented by the formula
\begin{eqnarray*}
\partial_t^2\phi(t,x)
&=&
(\partial_t^2\phi)_D +\oint_{|x-y|\leq t}
\int 
a^{\phi_{tt}}(\omega,p)f(t-|x-y|,y,p)\,dp \frac{dy}{|x-y|^3}\\
&&
{} +\int_{|x-y|\leq t}\int  b^{\phi_{tt}}
(\omega,p)(Sf)(t-|x-y|,y,p)\,dp \frac{dy}{|x-y|^2}\\
&&
{} +\int_{|x-y|\leq t}\int  c^{\phi_{tt}} 
(\omega,p) (S^2f)(t-|x-y|,y,p)\,dp \frac{dy}{|x-y|}\\
&=&
(\partial_t^2\phi)_D+(\partial_t^2\phi)_\mathrm{sing}+(\partial_t^2\phi)_S+
(\partial_t^2\phi)_{S^2},
\end{eqnarray*}
where $(\partial_t^2\phi)_D$ contains only integrals of derivatives of 
the initial data and the kernels are smooth.
Moreover, the kernel $a^{\phi_{tt}}$ of the most singular 
integral $(\partial_t^2\phi)_\mathrm{sing}$ has zero $\omega$-average:
\[
\int_{|\omega|=1}  a^{\phi_{tt}}(\omega,p)\,dS_\omega =0,\ p \in \R^3;
\] 
observe the principal value of the spatial integral in 
$(\partial_t^2\phi)_\mathrm{sing}$.
Similar representations with kernels having the same properties hold 
for the other second order derivatives of $\phi$. In particular, 
in each case the kernel of the most singular integral 
has zero $\omega$-average.
\end{Proposition}
\noindent\textit{Sketch of the proof:} 
The representation for $\partial_t^2\phi$ is obtained from (\ref{dtphirep1}) 
in the same way as Glassey and Strauss derive the representation for the
gradient of the electromagnetic field from (\ref{erep})--(\ref{brep}) 
cf.\ \cite[Thms.~4,5]{GS}. Each term in (\ref{dtphirep1}) is differentiated
with respect to $t$ and then (\ref{dtrep}) is used with $g=f$ and 
$g = Sf$. Integration by parts leads to the formula for
$\partial_t^2\phi$  after a straightforward calculation. 
The proof of the averaging property for $a^{\phi_{tt}}$
can be reduced to the case of \cite{GS} using the relation between 
the kernels for the Nordstr\"om-Vlasov system and those for the
relativistic Vlasov-Maxwell system
which we noted in Propositions~\ref{dtphirep} and \ref{dxphirep}.
The crucial point is that to get from $a^{E}$ and $a^{B}$
to $a^{\phi_t}$ or $a^{\phi_x}$ we need to multiply by
terms depending only on $p$. 
For example, in the case of $\partial_t^2\phi$ the most singular
integral is obtained by differentiating in time the third term in the 
right hand side of (\ref{dtphirep1}), substituting (\ref{dtrep}) and 
then integrating by parts in $y$. Hence the resulting
kernel is 
\begin{eqnarray*}
a^{\phi_{tt}}(\omega,p)
&=&
-|x-y|^3\,\widehat{p}\cdot\nabla_y\Bigg
[\frac{a^{\phi_t}(\omega,p)}{(1+\omega\cdot\widehat{p})|x-y|^2}
\Bigg]\\
&=&
-|x-y|^3\sqrt{1+p^2}\,\sum_{i,j}\widehat{p}_i\widehat{p}_j
\frac{\partial}{\partial y_j}\Bigg
[\frac{a_i^{E}(\omega,p)}{(1+\omega\cdot\widehat{p})|x-y|^2}\Bigg]\\
&=&
-|x-y|^3\sum_{i,j}\frac{\widehat{p}_i\widehat{p}_j}{\sqrt{1+p^2}}
\frac{\partial}{\partial y_j}\Bigg
[\frac{\omega_i+\widehat{p}_i}{(1+\omega\cdot\widehat{p})^3|x-y|^2}\Bigg]\\
&=&
\sqrt{1+p^2}\,\widehat{p}\cdot\tilde{a}(\omega,p),
\end{eqnarray*}
where $\tilde{a}(\omega,p)$ is the kernel which appears in (44) of \cite{GS}. 
Thus the averaging property for $a^{\phi_{tt}}$ 
follows from (43) of \cite{GS}. The analogous argument applies
to the other second order derivatives. \prfe  

\begin{Lemma} \label{ddphiest}
Under the assumption $\Lambda(T)<\infty$, a classical solution on
the time interval $[0,T[$ 
satisfies the estimate
\[
\|D^2 \phi(t)\|_\infty
\leq C_T\Big[1+\ln^*\sup_{0\leq\tau\leq t}\|Df(\tau)\|_\infty
+\int_0^t\|D^2 \phi(\tau)\|_\infty\,d\tau \Big],\ t\in [0,T[,
\]
where
\[
\ln^*s=\left\{\begin{array}{cl}
s&\textrm{ for $s\leq 1$}\\
1+\ln s &\textrm{ for $s\geq 1$}.\\
\end{array} \right.
\]
Combining this  with Corollary~\ref{dfest} and using Gronwall's inequality
implies
\[
\sup\{\|D^2 \phi (t)\|_\infty:\,0\leq t<T\}<\infty.
\]
\end{Lemma}
\noindent\textit{Sketch of the proof: }The proof is similar to
\cite[Lemma 6]{GS} and is outlined only for
$\partial^2_t\phi$. The data term in Proposition~\ref{ddtphirep} 
is estimated in terms of the appropriate norms 
of the initial data. The second term, i.e., the
most singular integral, is estimated using the averaging property
of its kernel as follows, cf.\ also   (59)--(63) of \cite{GS}:
\begin{eqnarray*}
&&
|(\partial_t^2\phi)_\mathrm{sing}(t,x)|
\leq
C_T \int_0^{t-d} (t-\tau)^{-1}\|f(\tau)\|_\infty d\tau\\
&&\qquad 
{} + \int_{t-d}^t \int_{|\omega| = 1}\int \left|a^{\phi_{tt}}(\omega,p)
\frac{f(t-\tau,x+\tau \omega,p) - f(t-\tau,x,p)}{\tau}\right|\,dp\,
dS_\omega\,d\tau\\
&&\quad \leq
C_T \ln(t/d) + C_T \sup_{0\leq\tau\leq t}\|D f(\tau)\|_\infty d
\end{eqnarray*}
which holds for any $0<d\leq t<T$. Let 
$\sigma = \sup_{0\leq\tau\leq t}\|D f(\tau)\|_\infty$. If $\sigma \leq 1$ choose
$d=t$, else choose $d=\min\{\sigma^{-1},t\}$. Then
\[
|(\partial_t^2\phi)_\mathrm{sing}(t,x)| \leq C_T
\left[1+\ln^*\sup_{0\leq\tau\leq t}\|D f(\tau)\|_\infty\right].
\]
For $(\partial_t^2\phi)_S$  we express $S f$ by (\ref{vlasov2}) and
integrate by parts in $p$. By the assumed bound on $\cal{P}$
the resulting kernels are bounded
on the support of $f$. Since the resulting 
expression contains only terms in $f$ and 
first order derivatives of $\phi$ which are bounded according to
Lemmas~\ref{fest} and \ref{dphiest} we obtain the bound on
$|(\partial_t^2\phi)_S|$.
To estimate $(\partial_t^2\phi)_{S^2}$ we expand $S^2f$ using the Vlasov 
equation:
\begin{eqnarray*}
S^2 f
&=&
4S[f\,S\phi] + S[((S\phi)\,p+
(1+p^2)^{-1/2}\nabla_x\phi)\cdot\nabla_p f]\\
&=&
4 f\,S^2 \phi +4 (S\phi)\,(S f) + (S^2\phi) p\cdot\nabla_p f +
(1+p^2)^{-1/2} (S\nabla_x\phi) \cdot\nabla_p f\\
&&
{}+F(t,x,p)\nabla_p(S f) - 
F_k \frac{\partial \widehat{p}_k}{\partial p_j}\partial_{x_j}f.
\end{eqnarray*}
After substituting this expression into $(\partial_t^2\phi)_{S^2}$ we 
remove the $p$-gradients by integration by parts.
Then we use (\ref{dyrep}) to express $\partial_{x_j}f$
and expand $Sf$ using the Vlasov equation. 
We end up with an expression which contains only first order
$p$-derivatives of $f$, 
second order derivatives of $\phi$ and the perfect derivative
$\nabla_y[f(t-|x-y|,y,p)]$.  
Integrating the corresponding terms by parts in $p$ 
and $y$ and using Lemmas~\ref{fest} and \ref{dphiest} again 
we get the estimate
\[
|(\partial_t^2\phi)_{S^2}(t,x)|\leq 
C_T\left[1+\int_0^t \|D^2 \phi(\tau)\|_\infty\,d\tau \right].
\]
Summing up the various estimates and using the same argument for all 
second order derivatives proves the first estimate of the lemma.
The second one then follows via Corollary~\ref{dfest}
and Gronwall's inequality. \prfe

\section{Proof of Theorem~\ref{locex}} \label{locexproof}
\setcounter{equation}{0}
Let us start by proving that the initial data (\ref{data})
with the regularity specified in the theorem launch a 
unique solution in some local time interval. For this purpose, we
define an approximation sequence $(f^{(n)}, \phi^{(n)})$ in the following 
way: For $n=0$ we set $f^{(0)}(t,x,p)=f^\mathrm{in}(x,p)$ and
$\phi^{(0)}(t,x)=\phi_0^\mathrm{in}(x)$. If
$(f^{(n-1)},\phi^{(n-1)})$ is already defined, we define the $n^\mathrm{th}$ 
iterate as the solution of the system
\[
S f^{(n)}-
[(S\phi^{(n-1)})p+(1+p^2)^{-1/2} \nabla_x\phi^{(n-1)}]\cdot
\nabla_p f^{(n)}
=4 f^{(n)}\, S\phi^{(n-1)} ,
\]
\[
\partial_t^2\phi^{(n)}-\Delta_x\phi^{(n)}=-\int
f^{(n)}\frac{dp}{\sqrt{1+p^2}},
\]
with initial data
\[
f^{(n)}(0,x,p)=f^\mathrm{in}(x,p),\quad
\phi^{(n)}(0,x)=\phi_0^\mathrm{in}(x), \quad
\partial_t\phi^{(n)}(0,x)=\phi_1^\mathrm{in}(x).
\]
We also define
\[
\mathcal{P}^{(n)}(t)=\sup
\left\{|p|: 0\leq s< t,\ (x,p) \in \supp f^{(n)}(s)\right\}
\]
and $f_{n,m}=f^{(n)}-f^{(m)},\,\phi_{n,m}=\phi^{(n)}-\phi^{(m)}$. 
Local existence of a classical solution will follow from
\begin{Proposition} \label{iterates}
For all $n\in \N$,
$f^{(n)}\in C^1([0,\infty[\times \R^6),\
\phi^{(n)}\in C^2([0,\infty[\times\R^3)$. There exists a time 
$T_*>0$ and a continuous function $z:[0,T_*[ \to [0,\infty[$ such that  
\[
\mathcal{P}^{(n)}(t)+\|f^{(n)}(t)\|_\infty+\|\phi^{(n)}(t)\|_\infty+
\|D \phi^{(n)}(t)\|_\infty\leq z(t),\ t\in[0,T_*[,\ n\in\N.
\]
Moreover
\[
\|f_{n,m}(t)\|_\infty + \|\phi_{n,m}(t)\|_\infty + 
\|D \phi_{n,m}(t)\|_\infty +
\|D f_{n,m}(t)\|_\infty + \|D^2\phi_{n,m}(t)\|_\infty\to 0,
\]
as $n,m\to\infty$, uniformly on any compact time
interval $[0,T]\subset[0,T_*[$.
\end{Proposition}
\noindent\textit{Proof: }We adapt the
estimates proved above for solutions of the Nordstr\"om-Vlasov system
to the approximation sequence $(f^{(n)},\phi^{(n)})$. 
First we notice that the functions $\phi^{(n)}$, 
$\partial_t\phi^{(n)}$ and $\nabla_x\phi^{(n)}$ have integral 
representations analogous to the ones proved for the solution:
\begin{equation} \label{phinrep} 
\phi^{(n)}(t,x)=\phi_0(t,x)-\int_{|x-y|\leq t}
\int f^{(n)}(t-|x-y|,y,p)\frac{dp}{\sqrt{1+p^2}}\frac{dy}{|x-y|},
\end{equation}
\begin{eqnarray*}
\partial_t\phi^{(n)}(t,x) 
&=&
\partial_t\phi_0(t,x)-t^{-1}
\int_{|x-y|=t}\int
\frac{f^\mathrm{in}(y,p)}{(1+\omega\cdot\widehat{p})}\frac{dp}{\sqrt{1+p^2}} 
dS_y \\
&&
{}- \int_{|x-y|\leq t}\int  
a^{\phi_t}(\omega,p)f^{(n)}(t-|x-y|,y,p)\,dp\frac{dy}{|x-y|^2}\\
&&
{}- \int_{|x-y|\leq t}\int  
b^{\phi_t}(\omega,p) [(S \phi^{(n-1)})f^{(n)}](t-|x-y|,y,p)\,dp\frac{dy}{|x-y|}\\
&&
{}- \int_{|x-y|\leq t}\int  
c^{\phi_t}(\omega,p)\cdot\nabla_x\phi^{(n-1)} 
f^{(n)}(t-|x-y|,y,p)\,dp\frac{dy}{|x-y|}
\end{eqnarray*}
and similarly for $\nabla_x\phi^{(n)}$. 
Moreover, if we denote by $(X_n,P_n)$ the solution of 
(\ref{xchar})--(\ref{chardata}) 
with $\phi$ replaced by $\phi^{(n-1)}$ then
we have, as in (\ref{frep1}) and (\ref{frep2}), 
\begin{eqnarray*}
f^{(n)}(t,x,p)
&=&f^\mathrm{in}(X_n(0),P_n(0))+4\int_0^t 
[f^{(n)} S \phi^{(n-1)}](\tau,X_n(\tau), P_n(\tau))\,d\tau ,\\
f^{(n)}(t,x,p)
&=&
f^\mathrm{in}(X_n(0),P_n(0))\exp \big \{4\phi^{(n-1)}(t,x)-
4\phi^\mathrm{in}(X_n(0)) \big\}.
\end{eqnarray*}
Thus the analogues of the estimates in Lemmas~\ref{phiest} and \ref{dphiest}
read
\begin{eqnarray*}
\|\phi^{(n)}(t)\|_\infty
&\leq& 
C_0(1+t) + c\,(1+t)\int_0^t 
\|f^{(n)}(\tau)\|_\infty \mathcal{P}^{(n)}(\tau)^{2}\,d\tau,\\
\|D \phi^{(n)} (t)\|_\infty
&\leq& 
C_1(1+t) \\
&&
{}+ c\,(1+t)\int_0^t  
(1+\|D \phi^{(n-1)} (\tau)\|_\infty)
\|f^{(n)}(\tau)\|_\infty\mathcal{P}^{(n)}(\tau)^{5}\,d\tau
\end{eqnarray*}
and hold for all $t\geq 0$. 
The analogue of the first estimate in Lemma~\ref{fest} is
\[
\|f^{(n)}(t)\|_\infty\leq \|f^\mathrm{in}\|_\infty+4\int_0^t  
\|D \phi^{(n-1)} (\tau)\|_\infty \|f^{(n)}(\tau)\|_\infty\,d\tau.
\]
Finally, as in Lemma~\ref{pest},
\[
\mathcal{P}^{(n)}(t)\leq P_0+2\int_0^t 
\|D \phi^{(n-1)} (\tau)\|_\infty(1+\mathcal{P}^{(n)}(\tau))\,d\tau .
\]
Let us define
\[
Q_n(t)=\mathcal{P}^{(n)}(t)+\|f^{(n)}(t)\|_\infty+
\|\phi^{(n)}(t)\|_\infty +\|D \phi^{(n)}(t)\|_\infty.
\]
Combining the previous estimates we obtain the inequality
\[
Q_n(t)\leq C+C
\int_0^t Q_{n-1}(\tau)Q_{n}(\tau)^6\,d\tau,\ t\in [0,1],
\] 
where 
\begin{equation} \label{cdef}
C=\max\{4,2C_0,2C_1\} P_0^4\left(1+P_0+\|f^\mathrm{in}\|_\infty+
\|\phi_0^\mathrm{in}\|_{2,\infty}+\|\phi_1^\mathrm{in}\|_{1,\infty}\right).
\end{equation}
It is then straightforward to show by induction that
\[
Q_n(t) \leq z(t),\ t\in [0,T_\ast[,\ n\in \N,
\]
where $z$ is the maximal solution of
\[
\dot{z} = C\, z^7,\quad z(0)=C, 
\]
which exists up to time $T_*=C^{-7}/6<1$. 
Since $\mathcal{P}^{(n)}$ is
bounded, the regularity of the approximation sequence follows by 
(\ref{phinrep}) and the induction hypothesis. 

In any time interval $[0,T]\subset
[0,T_*[$ and for any iterate, the quantity $\Lambda(T)$ is bounded by a 
constant $C_T$. By Corollary~\ref{dfest} and Lemma~\ref{ddphiest}
applied to the iterates we obtain uniform bounds on
the first order derivatives of $f^{(n)}$ and the second order derivatives
of $\phi^{(n)}$ on $[0,T]$. The Cauchy property for $f^{(n)}$
and for $\phi^{(n)}$ and its first order derivatives is then
easily deduced by deriving a Gronwall inequality for $f_{n,m}$. 
The Cauchy property for the next level of derivatives
is less simple, and we refer to the analogous
arguments for the relativistic Vlasov-Maxwell system, cf.\ 
\cite[pp.~82--85]{GS}. 
\prfe   

To prove uniqueness of local solutions
let $(f_1,\phi_1)$ and $(f_2,\phi_2)$ denote two solutions
with the same initial data which exist on some common
time interval $[0,T]$ and let $g=f_1-f_2$, $\psi=\phi_1-\phi_2$. 
For brevity we also let
$\lambda (t)=\|g(t)\|_\infty+\|D \psi(t)\|_\infty$.
Using the Vlasov equation we obtain
\begin{eqnarray*}
&&Sg - [(S\phi_1)p+ (1+p^2)^{-1/2} \nabla_x\phi_1]\cdot\nabla_p g\\
&&
\qquad\qquad=
4 g\, S\phi_1
+[(S\psi)\,p+(1+p^2)^{-1/2}\nabla_x\psi]\cdot\nabla_p f_2 + 4 f_2\, S\psi.
\end{eqnarray*}
The right hand side of this equation is estimated in modulus by 
$C_T\lambda (t)$. 
Hence integration along characteristics implies
\[
\|g(t)\|_\infty\leq C_{T}\int_0^t \lambda (s)\,ds.
\]
Taking the difference of the representation formulas for the 
derivatives of $\phi_1$ and $\phi_2$ and estimating as in 
Lemma~\ref{dphiest}
we obtain 
\[
\|D\psi(t)\|_\infty\leq C_{T}\int_0^t \lambda (s)\,ds.
\]
By the previous two estimates $\lambda =0$ on
$[0,T]$, and uniqueness is established. 

Assume now that the local solution has been extended to
a maximal time interval $[0,\tmax[$, and assume further that
$\Lambda(\tmax)<\infty$. To complete the proof of Theorem~\ref{locex}
we have to show that $\tmax = \infty$. Assume that $\tmax < \infty$.
Then due to Lemmas~\ref{fest}, \ref{phiest}, 
\ref{dphiest}, and \ref{ddphiest} the quantities
$\mathcal{P}(t)$, $\|f(t)\|_\infty$, $\|\phi(t)\|_\infty$, 
$\|D \phi(t)\|_\infty$, and $\|D^2 \phi(t)\|_\infty$ are uniformly 
bounded on $[0,\tmax[$. For any $t_0\in [0,\tmax[$ we 
consider the Cauchy problem for the Nordstr\"om-Vlasov system with 
$f(t_0),\,\phi(t_0),\, \partial_t\phi(t_0)$ 
prescribed as data at time $t_0$; let us ignore for the moment the
problem that these data do not have the regularity required in
our local existence result. 
By the same arguments as in Proposition~\ref{iterates} we conclude that this 
Cauchy problem has a local
classical solution in an interval $[t_0,t_0+\delta[$ where 
$\delta>0$ can be chosen independently of $t_0$. To see the latter observe
that all the quantities which entered the definition of the
local existence time $T_\ast$, cf.\ (\ref{cdef}), 
are now uniformly bounded on $[0,\tmax[$. 
Hence, if we take $t_0$ sufficiently close to $\tmax$, the solution 
$(f,\phi)$ can be extended beyond the time $\tmax$, 
which is a contradiction to its maximality.
Thus we have shown that the bound on $\Lambda$ implies that the
maximal solution is global.

To complete the proof let us deal with the technical catch that
$f(t_0),\,\phi(t_0),\, \partial_t\phi(t_0)$ do not have the regularity
required for our local existence result. Instead of using 
Proposition~\ref{iterates} directly we define a new sequence
of iterates as follows: For $(f^{(0)},\phi^{(0)})$ we take a global extension
of $(f,\phi)_{|[0,t_0]}$ with the same regularity as required of the solution
and with $\|f^{(0)}(t)\|_\infty$, $\|\phi^{(0)}(t)\|_\infty$,
$\|D \phi^{(0)}(t)\|_\infty$ uniformly bounded in $t$.
Given the $(n-1)^\mathrm{st}$ iterate we define the $n^\mathrm{th}$
iterate exactly as before. In particular, all these iterates coincide
with the solution on the time interval $[0,t_0]$, and the term $\phi_0$
in (\ref{kirchhoff}), which is the one in which the ``loss of derivative''
problem arises, is the one determined by the field data at $t=0$.
It is now straightforward to repeat all the estimates for these
new iterates and to show that they converge to a solution on
a time interval $[0,t_0 + \delta]$ where $\delta$ can be bounded away from
zero if the quantities entering the definition of $T_\ast$ in
Proposition~\ref{iterates} are bounded away from infinity, cf.\ (\ref{cdef}).
To put it shortly: Data obtained by evaluating a classical solution at some
time $t_0$ {\em are} admissible data for the local existence result.
\section{Proof of Theorem~\ref{blowup}} \label{blowupproof}
\setcounter{equation}{0}
The proof of Theorem~\ref{locex} is not 
affected by changing the sign of the right hand side of
(\ref{wave2}) and so the Cauchy problem 
for the ``repulsive'' Nordstr\"om-Vlasov system has a 
unique classical solution on a maximal
interval of time $[0,\tmax[$. 
Let us give data which are homogeneous in the ball $B_R(0)$ and 
assume that $\tmax>R$. Then the solution remains
homogeneous in the cone 
\[
\Omega(R)=\{(t,x):|x|\leq R-t,\,0\leq t\leq R\}. 
\]
This property can easily be understood as a consequence of finite 
propagation speed, but for the sake of completeness we give a proof
of this claim. We use the approximation sequence
which we define as in the proof of Theorem~\ref{locex}, but with the
sign of the second term in (\ref{phinrep}) reversed. 
The pair $(f^{(0)},\phi^{(0)})$
is homogeneous in $\Omega(R)$ by definition. Assume this is true at 
step $n-1$. If $(t,x)\in\Omega(R)$, then $(s,X_{n}(s))$ will
be contained in $\Omega(R)$ for all $s\leq t$. Hence
\[
\frac{dP_n}{ds}=-\partial_{s}\phi^{(n-1)}(s,0)P_n,\ (t,x)
\in\Omega(R),\,s\leq t.
\]   
Integrating this equation we obtain
\[
P_n(s)=p\,\exp\big (\phi^{(n-1)}(t,0)-\phi^{(n-1)}(s,0)\big )
\]
and so $P_n(0)=p\,\exp\big (\phi^{(n-1)}(t,0)-\phi^\mathrm{in}_0(0)\big )$. 
Hence
\[
f^{(n)}(t,x,p)=f^\mathrm{in}(0,P_n(0))\exp(4\phi^{(n-1)}(t,0)-
4\phi_0^\mathrm{in}(0)),
\ (t,x)\in\Omega(R)
\]
and so $f^{(n)}$ is homogeneous on $\Omega(R)$. By the Kirchhoff formula
(\ref{phinrep}) with the plus sign between the two terms this 
is true for $\phi^{(n)}$ as well. Since the solution $(f,\phi)$
arises as a uniform limit of the iterative sequence on some time interval
we conclude that it is homogeneous in the truncated cone 
$\Omega_T(R)=\{(t,x):|x|\leq R-s,\,0\leq s\leq T\}$ for some
$0<T\leq R$. Choose $\tilde{T} \leq R$ maximal such that
the solution is homogeneous on $\Omega_{\tilde{T}}(R)$. 
If $\tilde{T}<R$ then we could construct the approximation sequence with data 
$f(\tilde{T})$, $\phi(\tilde{T})$, $\partial_t \phi(\tilde{T})$
at time $\tilde{T}$, which are homogeneous on the ball $|x|\leq
R-\tilde{T}$, and we would be able to prove homogeneity of the solution in 
the truncated cone $\Omega_{\tilde{T}+\delta}(R)$ for some
$\delta>0$ which is a contradiction to the maximality of $\tilde{T}$. 

Hence the solution $(f,\phi)$ is homogeneous on the
cone $\Omega_R(R)=\Omega (R)$. If by abuse of notation we denote this
spatially homogeneous solution on $\Omega (R)$ by $f=f(t,p),\ \phi=\phi(t)$
then this pair satisfies the equations
\begin{equation} \label{homwave}
\ddot\phi=\mu,
\end{equation}
\begin{equation} \label{hommu}
\mu(t) = \int f(t,p)\frac{dp}{\sqrt{1+p^2}}
\end{equation}
\begin{equation} \label{homvlasov}
\partial_t f-\dot\phi\, p\cdot\nabla_p f=4\dot\phi f,
\end{equation}  
for all $t\in[0,R]$, where $\cdot=d/dt$. 
Hence $\ddot\phi \geq 0$, 
and since $\dot\phi(0)>0$, $\phi$ is increasing.
Moreover
\[
\dot\mu=\dot \phi\;\int (p\cdot\nabla_p f+4 f)\frac{dp}{\sqrt{1+p^2}}
=\dot\phi \;\int  f\frac{1+2p^2}{(1+p^2)^{3/2}}\,dp \geq \dot\phi\mu.
\]
Integrating this inequality we 
obtain $\mu(t)\geq \mu(0) e^{\phi(t)-\phi(0)}$.
Hence by (\ref{homwave}),
$\ddot\phi\geq \mu(0) e^{\phi-\phi(0)}$, and
\[
\frac{d}{dt}\left[(\dot\phi)^2- 2\mu(0) e^{-\phi(0)}e^{\phi}\right]\geq 0.
\]
By our assumptions in Theorem~\ref{blowup} the term in square brackets  
is non-negative at time $t=0$. Hence
$(\dot\phi(t))^2-2\mu(0) e^{-\phi(0)}e^{\phi(t)}\geq 0$ for all
$t\in[0,R]$. Setting 
$\lambda=2\mu(0) e^{-\phi(0)}$, the latter inequality entails
\[
-2\frac{d}{dt}(e^{-\phi/2})\geq\sqrt{\lambda},
\]  
which implies 
\[
e^{-\phi(t)/2}\leq e^{-\phi(0)/2}-\frac{1}{2}\sqrt{\lambda}\,t,\
t\in[0,R].
\]
Since by our assumptions in Theorem~\ref{blowup} the
right hand side of this inequality is not positive for $t=R$
this is a contradiction. Hence the length of the maximal existence interval 
cannot be larger that $R$, and the proof of 
Theorem~\ref{blowup} is complete.

\section{The one dimensional case} \label{1d}
\setcounter{equation}{0}
In this final section we study the one dimensional system
(\ref{wave1d}), (\ref{mudef1d}), (\ref{vlasov1d}), in particular, $x, p \in \R$
and all integrals are one dimensional.
We integrate the one dimensional wave equation (\ref{wave1d})
along its characteristics to obtain
\begin{eqnarray}
\partial_t \phi(t,x)
&=&
\frac{1}{2} \left[ (\partial_t \phi + \partial_x \phi)(0,x+t) +
(\partial_t \phi - \partial_x \phi)(0,x-t) \right]\nonumber \\
&&
{}- \frac{1}{2} \int_0^t\left[\mu(s,x+s-t) -\mu(s,x-s+t)\right]\, ds,
\label{dtphirep1d}\\
\partial_x \phi(t,x)
&=&
\frac{1}{2} \left[ (\partial_t \phi + \partial_x \phi)(0,x+t) -
(\partial_t \phi - \partial_x \phi)(0,x-t) \right]\nonumber \\
&&
{}+ \frac{1}{2} \int_0^t\left[\mu(s,x+s-t) -\mu(s,x-s+t)\right]\, ds.
\label{dxphirep1d}\end{eqnarray}
Using these and the resulting formulas for $\phi$ itself and its
second order derivatives instead of the representation formulas
in Section~\ref{fieldrep} it is straight forward to show that initial
data of the regularity stated in Theorem~\ref{globex1d} launch a
classical solution on some maximal time interval $[0,\tmax[$,
and $\tmax = \infty$ if $\phi$ and the momenta in the support of $f$
remain bounded, i.e., the analogue of Theorem~1 holds.
The only point that needs to be checked here is that a bound on
$\mathcal{P}(t)$, $\|f(t)\|_\infty$, $\|\phi(t)\|_\infty$, and
$\|D \phi(t)\|_\infty$ implies a bound on the second order derivatives
of $\phi$, i.e., the analogue of Lemma~\ref{ddphiest} holds. The crucial
terms to estimate if one for example takes another time derivative in
(\ref{dtphirep1d}) are of the form
\begin{equation} \label{dd1d}
\int_0^t\int (\partial_x f)(s,x+s-t,p) \frac{dp}{\sqrt{1+p^2}}
\end{equation}
To get rid of the derivative on $f$ we use the relation
\begin{eqnarray*}
(\partial_x f)(s,x+s-t,p) 
= 
\frac{1}{1+\widehat{p}} 
&\Bigl[&\partial_p\left([(S\phi) p + (1+p^2)^{-1/2} \partial_x \phi]f\right)
(s,x+s-t,p) \\
&&
{} + (S\phi f)(s,x+s-t,p) - \partial_s[f(s,x+s-t,p)]\Bigr]
\end{eqnarray*}
which follows from the Vlasov equation (\ref{vlasov1d}).
Upon substitution into (\ref{dd1d}) we can integrate by parts with respect to
$p$ and $s$ respectively. The resulting terms will contain only first order
derivatives of $\phi$, no derivatives of $f$, and kernels which are
bounded as long as $p$ is bounded away from infinity.
 
Assume now that $\tmax < \infty$. Then conservation of 
energy implies that $\phi$ remains bounded on the $x$-support
of $f$. To see this, note that for $R_0>0$
sufficiently large,
$f(t,x,p)=0$ for $|x|>R_0 + t$ , and hence we can change $\phi$
outside this forward lightcone without affecting the solution
inside. Together with the bound on the spatial integral of 
$\partial_x \phi$ coming from conservation of energy this implies
a bound on $\phi$ on the $x$-support of $f$.
By (\ref{frep2}) this in turn implies a bound on $f$ as well;
$e^{-2\phi} f$ is constant along characteristics, cf. (\ref{vlasov3n}).
Therefore,
\[
\|\mu(t)\|_\infty \leq 2 \|f(t)\|_\infty \int_0^{\mathcal{P}(t)}
\frac{dp}{\sqrt{1+p^2}}
\leq C \ln(1+\mathcal{P}(t))
\]
with $C>0$ independent of $t$. Combining this with the representation
formulas (\ref{dtphirep1d}) and (\ref{dxphirep1d}) implies
\[
\|D \phi(t)\|_\infty \leq C \left[1+\ln(1+\mathcal{P}(t))\right].
\]
Using this estimate in (\ref{pchar}) of the characteristic system
and integrating we obtain the inequality
\[
\mathcal{P}(t) \leq P_0 + C \int_0^t(1+\mathcal{P}(\tau))
\ln(1+\mathcal{P}(\tau))\, d\tau
\]
which holds for all $t\in[0,\tmax[$ with some constant $C>0$ which
does not depend on $t$. Since by assumption $\tmax < \infty$ 
this implies that $\mathcal{P}$ is bounded as well which is a contradiction.
Hence $\tmax = \infty$.  

\bigskip
\noindent
{\bf Acknowledgments:}
S.~C.\ acknowledges support by the European HYKE network 
(contract HPRN-CT-2002-00282),
G.~R.\ acknowledges support by the Wittgenstein 
2000 Award of P.~A.~Markowich.


\end{document}